\documentclass{article}

 \usepackage[final,nonatbib]{nips_2016}
\usepackage{multicol}
\usepackage[pdftex]{graphicx}
\usepackage{multirow}
\usepackage{verbatim}
\usepackage{booktabs,mathptmx,siunitx}
\graphicspath{{figures/}}
\DeclareGraphicsExtensions{.pdf,.jpeg,.png}
\usepackage[ruled]{algorithm2e}
\SetAlFnt{\small}

\usepackage{color}

\usepackage[utf8]{inputenc} 
\usepackage[T1]{fontenc}    
\usepackage{hyperref}       
\usepackage{url}            
\usepackage{booktabs}       
\usepackage{amsfonts}       
\usepackage{nicefrac}       
\usepackage{microtype}      
\usepackage[caption=false,font=footnotesize]{subfig}

\title{\textit{dMath}: Distributed Linear Algebra for DL}

\author{
Steven Eliuk, Cameron Upright, Hars Vardhan, Stephen Walsh, Trevor Gale\\
Samsung Electronics\\ 
Computing Science Innovation Center, SRA-SV\\
665 Clyde Avenue\\
Mountain View, CA 94043\\
Email: \{s.eliuk,c.upright\}@samsung.com\\
}

\begin{document}

\maketitle

\begin{abstract}
The paper presents a parallel math library, \textit{dMath},
that demonstrates leading scaling when using intranode, internode, and hybrid-parallelism
for deep learning (DL).  \textit{dMath} provides easy-to-use distributed primitives and a
variety of domain-specific algorithms including matrix multiplication,
convolutions, and others allowing for rapid development of 
scalable applications like deep neural networks (DNNs).
Persistent data stored in GPU memory and advanced memory management techniques
avoid costly transfers between host and device.
\textit{dMath} delivers performance, portability,
and productivity to its specific domain of support.  

\end{abstract}

\section{Introduction}
High-speed machine learning is becoming one of the most important areas of 
high performance computing (HPC) in the commercial, academic, and other spaces.
Machine-learning algorithms leverage traditional scientific
computing -- correlations, convolutions, FFTs, matrix and tensor
multiplications, and combinations thereof.  Thus, central to the solution
of key machine learning algorithms 
is the need  for both scalable architectures and
algorithmic libraries that implement these kernels efficiently with 
strong 
scaling.
This paper presents \textit{dMath}, a new scalable distributed math library.  {\em dMath} provides key linear algebra operations, convolutions and other fundamental algorithms  in the implementation of deep neural networks (DNNs).  
The key features of the library include:
a) support for persistent storage of operands in GPU memory; 
b) a data management service to cache shared objects over the course of data parallel operations;
c) data reorganization to support optimization of operations in series; 
d) a mixed precision mode, e.g. double, float, and half;
e) an automatically tuned, extendable data loading and augmentation pipeline;
f) an effective master-worker model to allow users to utilize {\em dMath} without requiring detailed knowledge of CUDA, MPI, or data distribution.  
The library uses pooling of unused GPU memory to avoid costly CUDA memory allocations and registrations with the InfiniBand (IB) driver.

Another notable feature of the system is the ability to ``keep what
you've seen.''  Because the data management layer has semantic
understanding of matrices and vectors, as algorithms
progress, portions of the parallel matrix can be retained in a cache
(within each MPI process). This allows for reduced communication in subsequent
steps, such as in the back-propagation stage of DNN training. 
For problems where computation is memory bound, caching is avoided.

Lastly, we exploit GPU-enabled MPI to enhance performance and utilize 
non-blocking MPI operations to  overlap communication
and computation, where appropriate.   

\nocite{GDR-reference1,GDR-reference2}
\section{\textit{dMath} Architecture}
\label{sec:dmath-architecture}

\textit{dMath} was designed based on a number of requirements:
a) support key algorithms for DNN pipelines;
b) utilize multiple GPUs together with multiple MPI processes
to reduce time to solution; 
c) exploit the latest GPUs from NVIDIA, the latest InfiniBand from Mellanox,
and the latest MPIs that utilize GDR in order to maximize distributed performance;
d) exploit the availability of PCI switching in order to gain density within
x86-64 servers;
and 
e) support basic fault tolerance through checkpoint-restart.

\textit{dMath} runs as a set of MPI processes in master-slave paradigm. 
It provides high-level abstraction for managing and operating on distributed data.
The developer uses \textit{dMath} like any
other mathematics library; the distributed computation is handled
internally, and the 
user is not required to have knowledge of the distributed multi-GPU implementation.



\subsection{Managing Persistent Data in the GPUs}
When working with GPU accelerators, copying data across the CPU-GPU boundary is extremely undesirable. Persistent storage of operands within the GPUs is therefore a critical feature of \textit{dMath} that improves the efficiency of CUDA-accelerated computation. Strong adherence to this model means that comparatively little of the multicore CPUs are utilized apart from servicing MPI middleware APIs and underlying MPI transport.




In \textit{dMath}, a distributed matrix is split into multiple non-overlapping
blocks that are stored on individual workers. Each worker is aware of
the layout of every matrix, which allows the workers to operate without the 
intervention of the master process.
%
%
%
Often it is desirable to have a copy of a matrix on each worker. In situations where the data rarely changes and memory is abundant, this sort of caching is beneficial to reduce communication. 
\textit{dMath} supports both synchronous and asynchronous matrix replication.


Replication allows us to perform efficient parameter redistribution when training deep neural networks. 
After each worker computes the weight updates for its chunk of the model, asynchronous replications are initiated for learnable parameters that will be needed by all workers for the forward pass.
This effectively overlaps parameter updates with the forward pass computation.

\subsection{Data Loading}
%

As training accelerates, larger data, and more complex data augmentation
pipelines can become the bottleneck when training in many-GPU
distributed systems. To avoid this bottleneck, data augmentation is done
in parallel with network training. dMath supports multi-threading, and
the movement of the computation for individual stages in the the data
augmentation pipeline between host and device. At runtime, dMath
dynamically tunes the number of worker threads and the location of each
data augmentation operation to optimize overall iteration time. Host
computation, transfers to the device, and computation on the GPU are overlapped
to reduce the processing time for each datum. To minimize data transfers
from the host to the device, promotion of data to higher precision types is done
lazily by the pipeline as needed for higher precision operations (e.g.,
mean subtraction).

\subsection{Reproducibility}
The ability to reproduce results is incredibly important, and when
certain subroutines are stochastic in nature, one can get different
results that may be difficult to duplicate. In \textit{dMath}, we use
seed values that are distributed via the master node to workers to
ensure reproducible results. However, there are a few subroutines
where concurrency and non-deterministic ordering of operations
can lead to small differences in results. For example, in the
distributed version of our \textit{AddRowColSumMatrix} subroutine we
sacrifice deterministic outcomes for speed and scalability.

\section{\textit{dMath} Kernels}
\label{sec:Algorithms}
\textit{dMath} combines a set of innovative ideas that simplify writing parallel
algorithms while supporting key kernels. In this section, we consider some of the key 
features of \textit{dMath} operations.

\subsection{Algorithms}
\textit{dMath} provides many numerical kernels/operations that are distributed over MPI, including
hundreds of algorithms and methods common to a DNN computational
pipeline.
These building blocks allow a varierty of DNN pipelines to be created (implicitly with distributed backends).

\subsection{Data Distribution Independence}
Algorithms in \textit{dMath} are correct independent of the how the distributed objects are
mapped to the workers.
\textit{dMath} performs any needed communication to ensure compatibility, rather 
than limiting the distributions to block-cyclic (and/or linear) 1D or 2D decompositions.
Previous libraries ({\em e.g.,} \cite{van1990data,skjellum-toolbox}) achieve data
distribution independence and varying degrees of compatibility with rectangular matrices
and cartesian decompositions, but notably require that the objects be laid out compatibly at the beginning of the GEMM function, rather than offering remapping services. As with other 
libraries, the shape of the data and concurrency 
can affect
the performance of \textit{dMath} kernels.

\subsection{Data Reorganization and Caching}
Where appropriate, \textit{dMath} allows an algorithm to
reshape (including a change of concurrency and layout), over the same group of processes 
or a superset/subset, and/or
change precision during reshape.
All metadata for a distributed operation can also be cached.
We utilize this feature when dealing with fixed pipelines and it provides greater scalability 
by allowing the workers to remember the entire forward and backward computations. This prevents thousands of costly broadcasts of metadata to the workers by replacing each one with a single 
cached identifier.



\section{DNN Training Results}
\label{sec:examples}



In this section, we present performance results training AlexNet~\cite{NIPS2012-4824}, 
GoogLeNet v1~\cite{szegedy-going-deeper}, and Inception v3~\cite{szegedy-rethinking-inception} 
using Expresso, a Samsung internal fork of Caffe, powered by \textit{dMath}, that employs 
hybrid parallelism~\cite{Krizhevsky14}. 
We compare Expresso to leading open-source frameworks.


\subsection{Scaling Experiments}
In the left-hand side of Table \ref{table:frameworkComparisons} we provide scaling 
performance results for Expresso v0.5. We compare our performance to Nvidia's 
NVcaffe v0.14~\cite{nvidia_multi12}, which provides leading open-source intranode 
scaling via data parallelism. The AlexNet experiments use a batch size of 1024 on
2-64 GPUs, and a 512 batch size on 1 GPU due to memory constraints. All tests in
Table \ref{table:frameworkComparisons} were conducted on Nvidia Tesla K80 GPUs.
Experiments with GoogLeNet were conducted using a 1024 batch size on 8-64 GPUs, and 128
batch size for 1-4 GPUs. 

On startup, \textit{dMath} automatically selects the optimal convolution algorithm
based on timing samples and system constraints. The single asterisk on \textit{Expresso}
results designates a sub-optimal algorithm choice because of memory constraints; hence
the super-linear scaling from two to four GPUs. The double asterisk signifies
GPU memory thrashing, which is a significant detriment to the performance of NVcaffe
for some batch sizes.

While it is significantly easier to weakly scale training, it eventually leads
to large batch sizes that could potentially harm the convergence of the network 
parameters. The results from these experiments demonstrate weak scaling from 
1-2 GPUs and 4-8 GPUs for AlexNet and GoogLeNet respectively, but maintain the
batch size beyond these points.

\begin{table}
\caption{Framework Comparison} \label{table:frameworkComparisons}
\scriptsize
\begin{center}
\vspace{-0.1in}
\begin{tabular}{|c|c|c|c|c||c|c|c|c|}
\hline
Number  & \multicolumn{2}{c|}{AlexNet} & \multicolumn{2}{c||}{GoogLeNet v1}  & \multicolumn{4}{c|}{AlexNet} \\
\multirow{2}{*}{of} &       \multicolumn{2}{c|}{1024 Batch}     & \multicolumn{2}{c||}{1024 Batch} &       \multicolumn{4}{c|}{256 Batch}\\
 &       \multicolumn{2}{c|}{ (FPS)}     & \multicolumn{2}{c||}{(FPS)}   &       \multicolumn{4}{c|}{ (FPS)}\\
GPUs &           Expresso  &    nv-caffe  & Expresso & nv-caffe &  Expresso &    CNTK  & CNTK (1-bit) & nv-caffe \\
 & v0.5 & v0.14   & v0.5 & v0.14 & v0.5 &  r2016-02-08 &  r2016-02-08 & v0.14 \\
\hline
1        & \textbf{479}      &     413   & \textbf{115} & 102 & 533      &     580   & \textbf{568} & 350\\
\hline
2       &   \textbf{*940} &  **682 & \textbf{215} & 205 &   \textbf{915} &  487 & 485  & 711\\
4       &  \textbf{1996} &  1165 &  \textbf{370} & 341  &  \textbf{1440} &  428 & 416 & 898\\
8       &  \textbf{3103} &  2204 &  \textbf{873} & **510 &  \textbf{1702} &  - & -  & 970 \\
16      & \textbf{4198} & 2615 &  1498 & \textbf{1515} & \textbf{2008} & - &  - &  875\\
32      & \textbf{5187} & \textbf{-} & \textbf{2330}  & \textbf{-} & \textbf{2104} & - & -  & - \\
64      & \textbf{5786}  & \textbf{-} & \textbf{3025} & \textbf{-}  & \textbf{2271}  & - & - & - \\
\hline
Mem. GB (16 GPUs) &  \textbf{2.29}  & 2.54  & \textbf{5.27} & 7.65  & - & - & - & - \\
\hline
Accuracy(Top-1\%)    &   \textbf{55.38} &  55.14 & \textbf{65.39} & 64.96  &   \textbf{58.59} &  - & - & 57.01\\
\hline 
\end{tabular}
\end{center}
\end{table}

\textit{Expresso} provides class-leading performance for strong scaling, as shown in the right-hand side of Table~\ref{table:frameworkComparisons}. 
It is one of \textit{dMath}'s fundamental goals to provide the ability to experiment 
with extremely large models that are stored in device memory, without being constrained 
to a subsets of GPUs because of poor strong scaling. The right-hand side of Table~\ref{table:frameworkComparisons} shows performance results for \textit{Expresso}, 
Microsoft's CNTK, and NVcaffe, scaling a 256 sample batch from 1-64 GPUs. Testing of CNTK 
was performed with both the regular and one-bit quantized versions of SGD. Across all
experiments, \textit{Expresso} provides superior scaling while also achieving higher accuracy and
a lower memory footprint due to its hybrid data/model parallelism scheme.
After extensive debugging, we were not able to successfully run the multi-GPU version of 
CNTK for anything but a few iterations and cannot provide accuracy metrics.

\subsection{Accuracy \& Half-Precision}
Accuracy is an extremely important metric and it is an
important attribute of \textit{dMath}. \textit{dMath} solves the harder problem, that of hybrid parallelism, and accuracy is never impacted regardless of the number of GPUs. 
Reduced precision data types enable even better scaling in a
distributed environment by reducing data transfer size.
\textit{dMath} supports half-precision for storage and computation. 
When using GPUs that do not support half-precision computation, 
\textit{dMath} works in a mixed-mode in which values are stored in half and upcast to float before 
computation.
With half-precision integration in \textit{Expresso}, we can train and test larger network on 
the same devices.
Preliminary results indicate that
\textit{Expresso} performs at par in mixed half-mode for inference accuracy on trained 
GoogLeNet and AlexNet networks.
Moving further, we are currently investigating the performance of training and testing on devices with true half-precision support. 

\subsection{Updated Experimental Results}
Using \textit{Expresso} v0.7 on 32 Tesla K80s ({\em i.e.,} 64 GPUs) with a 32 batch size per GPU, we are able to train Google's Inception v3 model at 1443 FPS. Training the IV3 model with a simplistic data augmentation pipeline (random cropping and mirroring) produced 92.8\% top-5 and 75.1\% top-1 single-crop test accuracy after 200 epochs, with a total runtime of 49.3hrs.
Consider 
Google's internal TensorFlow takes 65hrs on 100 Tesla K40 GPUs~\cite{google-tf-r0.8}, and it is quickly apparent that more 
HPC centric approaches provide superior performance. 
Our own experiments with open-source TensorFlow r0.8 were 
unable to get over 200 FPS on 8, 16, 32, or 64 Tesla K80s, or over 150 FPS on 4 or 8 Tesla M40s. Our TensorFlow experiments used a single parameter server and a 32 batch size per GPU.
Scaling out training with Expresso to 96 Tesla M40s with an 80 batch size per GPU, 
we are able to further reduce the training time to 14hrs with a frame rate of 5150 FPS. 
\section{Conclusion \& Future Work}
\label{sec:conclusion}
In this paper we presented \textit{dMath}, a scalable parallel math library that provides a complete set of primitives for DNN pipelines.
We showed experimental results through \textit{Expresso}, demonstrating superior scaling to leading open-source frameworks.
%
We have demonstrated the effectiveness of \textit{dMath}, but we also have a version of the popular open-source speech recognition library Kaldi~\cite{PoveyASRU2011},
powered by \textit{dMath}, that shows the generality of the library. In the future we wil be introducing new abstractions for alternative accelerators and communication devices, providing greater degrees of asynchronous behavior to increase scalability, and will be exploring ways to further reduce the memory footprint.

\newpage
\section*{Acknowledgment}
The authors 
thank all research collaboration partners,
internal and external to Samsung Electronics Company. Specifically, 
we 
acknowledge Rolf VandeVaart, from Nvidia, for his 
help with OpenMPI, the Ohio State University MVAPICH group, Nvidia, 
Cirrascale, and Mellanox.


\medskip

\small
\bibliographystyle{IEEEtran}

\end{document}